\documentclass[numbers,preprint,review,11pt]{elsarticle}
\biboptions{square,comma,sort&compress}
\usepackage[left=1in,right=1in,bottom=1in,top=1in]{geometry}
\usepackage{bm}

\usepackage{charter}
\usepackage[T1]{fontenc}
\usepackage{ae,aecompl}

\usepackage{amssymb,amscd}
\usepackage{url}
\usepackage{amsthm,amsfonts,amsmath,amsxtra}
\usepackage{graphicx}
\usepackage{float} 
\usepackage{xcolor}

\usepackage{graphicx}       
\usepackage{amssymb}
\usepackage{amsthm}
\usepackage{float}
\usepackage{rotating}

\usepackage{soul}

\usepackage{rotating}
\usepackage{amsmath,amsfonts,latexsym}
\usepackage[utf8]{inputenc}
\usepackage{mathrsfs}
\usepackage{color,subfig}
\usepackage{xcolor}
\usepackage{caption}
\usepackage{lscape}
\graphicspath{{./figures/}{otherfolder}}
\usepackage{array}
\usepackage{amstext}
\usepackage{placeins}
\newcommand*\patchAmsMathEnvironmentForLineno[1]{%
	\expandafter\let\csname old#1\expandafter\endcsname\csname #1\endcsname
	\expandafter\let\csname oldend#1\expandafter\endcsname\csname end#1\endcsname
	\renewenvironment{#1}%
	{\linenomath\csname old#1\endcsname}%
	{\csname oldend#1\endcsname\endlinenomath}}%
\newcommand*\patchBothAmsMathEnvironmentsForLineno[1]{%
	\patchAmsMathEnvironmentForLineno{#1}%
	\patchAmsMathEnvironmentForLineno{#1*}}%
\AtBeginDocument{%
	\patchBothAmsMathEnvironmentsForLineno{equation}%
	\patchBothAmsMathEnvironmentsForLineno{align}%
	\patchBothAmsMathEnvironmentsForLineno{flalign}%
	\patchBothAmsMathEnvironmentsForLineno{alignat}%
	\patchBothAmsMathEnvironmentsForLineno{gather}%
	\patchBothAmsMathEnvironmentsForLineno{multline}%
}

\newcommand{\cH}{\mathcal{H}}

\newcommand{\cP}{\mathcal{P}}

\makeatletter
\providecommand\phantomcaption{\caption@refstepcounter\@captype}
\makeatother
\usepackage{setspace}
\usepackage{multirow}
\usepackage{chngcntr}
\usepackage[normalem]{ulem}
\usepackage{relsize}

\usepackage{lineno}

\setcitestyle{aysep={}}

\journal{}

\makeatletter
\renewcommand\section{\@startsection {section}{1}{\z@}%
	{-4.5ex \@plus -1ex \@minus -.2ex}%
	{2.3ex \@plus.2ex}%
	{\centering\normalfont\Large}}
\renewcommand\subsection{\@startsection{subsection}{2}{\z@}%
	{-4.5ex \@plus -1ex \@minus -.2ex}%
	{2.0ex \@plus.1ex}
	{\large \bf}}
\makeatother

\makeatletter
\def\ps@pprintTitle{%
	\let\@oddhead\@empty
	\let\@evenhead\@empty
	\def\@oddfoot{}%
	\let\@evenfoot\@oddfoot}
\makeatother

\DeclareMathAlphabet{\mathscrbf}{OMS}{mdugm}{b}{n}

\makeatletter
\long\def\pprintMaketitle{\clearpage
	\iflongmktitle\if@twocolumn\let\columnwidth=\textwidth\fi\fi
	\resetTitleCounters
	\def\baselinestretch{1}%
	\printFirstPageNotes
	\begin{center}%
		\thispagestyle{pprintTitle}%
		\def\baselinestretch{1}%
		\Large\@title\par\vskip18pt
		\normalsize\elsauthors\par\vskip10pt
		\footnotesize\itshape\elsaddress\par\vskip36pt
	\end{center}%
	\gdef\thefootnote{\arabic{footnote}}%
}
\makeatother

\begin{document}
	
	\begin{frontmatter} 
		
		
		
		\title{\vspace*{250pt}
			Stochastic processes and host-parasite coevolution: linking coevolutionary dynamics and DNA polymorphism data
		}
		
		\author[lab1]{Wolfgang Stephan}
		\author[lab2]{Aur\'elien Tellier}
		
		\address[lab1]{Leibniz Institute for Evolution and Biodiversity Science, Natural History Museum, Berlin, Germany} 
		\address[lab2]{Professorship for Population Genetics, Department of Life Science Systems, School of Life Sciences,
			Technical University of Munich, Freising, Germany} 
	
	\end{frontmatter} 
	
	\setcounter{page}{0}
	
	
	\newpage

	\subsection*{Abstract}
Between-species coevolution, and in particular antagonistic host-parasite coevolution, is a major process shaping within-species diversity. In this paper we investigate the role of various stochastic processes affecting the outcome of the deterministic coevolutionary models. Specifically, we assess 1) the impact of genetic drift and mutation on the maintenance of polymorphism at the interacting loci, and 2) the change in neutral allele frequencies across the genome of both coevolving species due to co-demographic population size changes. We find that genetic drift decreases the likelihood to observe classic balancing selection signatures, and that for most 
realistic values of the coevolutionary parameters, balancing selection signatures cannot be seen at the host loci. Further, we reveal that contrary to classic expectations, fast changes in parasite population size due to eco-evo feedbacks can be tracked by the allelic site-frequency spectrum measured at several time points. Changes in host population size are, however, less pronounced and thus not observable. Finally, we also review several understudied stochastic processes occurring in host-parasite coevolution which are of importance to predict maintenance of polymorphism at the underlying loci and the genome-wide nucleotide diversity of host and parasite populations.

\newpage
\section{Introduction}
Antagonistic coevolution \index{coevolution} between hosts and parasites is defined as the reciprocal changes in allele frequencies in each species over time. Host-parasite coevolution is classically studied using deterministic dynamical system \index{dynamical system} models of equations for host and parasite populations. The underlying principle is that coevolution is driven by so-called negative indirect frequency-dependent selection (niFDS, \cite{WSAT_TellierandBrown2007a}) between host and parasite, an effect often referred to as the advantage of the rare. In other words, an allele in low frequency in the host population provides improved resistance and thus has an advantage. This allele is selected and increases in frequency. As it increases in frequency a corresponding matching allele in the parasite becomes selected, and thus the advantage of the host allele decreases with its increasing frequency \cite{WSAT_MayandAnderson1983, WSAT_TellierandBrown2007a}. \\

Frequency-dependent selection \index{selection!frequency-dependent} in coevolutionary models yields two extreme dynamics called arms race and trench warfare. The arms race dynamic is defined by the recurrent fixation of traits or alleles (\textit{e.g.} resistance in hosts and virulence in parasites) in both antagonistic species, while the trench warfare is defined by the maintenance over time of several trait values or alleles in each species. Based on deterministic models of coevolution, these two dynamics can be distinguished because they generate different expectations regarding trait evolution and/or allelic diversity over time \cite{WSAT_Holub2001,WSAT_gandon2008}. With the advance of DNA sequencing technologies, it is now possible to search for the loci underpinning coevolution in host and parasite genomes. Arms race and trench warfare are indeed classically expected to generate different signatures in polymorphism data (Single Nucleotide Polymorphisms, SNPs) obtained from several sequenced individuals per species \cite{WSAT_Woolhouseetal2002}. On the one hand, arms race dynamics are expected to exhibit signatures of selective sweeps \index{selection!positive} due to rapid recurrent fixation of alleles, such as low nucleotide diversity and skew of the site-frequency spectrum \index{site-frequency spectrum} (SFS) towards an excess of rare and high frequency variants (SNPs). On the other hand, trench warfare dynamics are expected to result in balancing selection \index{selection!balancing} signatures with high nucleotide diversity and an SFS with an excess of intermediate frequency variants (SNPs) \cite{WSAT_Woolhouseetal2002}. These aforementioned excess signatures of SNPs are defined compared to a neutral SFS in host and parasite populations. \\

However, in order for such genome-wide selection scans for genes under coevolution to be accurate, two questions need to be answered.
1) Can we disentangle signatures of coevolution from the variance in polymorphism signatures across the genome?
There is thus a need to model the variance in polymorphism data across the genomes of hosts and parasites, for example using the Kingman coalescent \index{coalescent} (or other coalescent models) with changes in population size. The past demographic history of the population is indeed a well-known confounding effect reducing the ability to detect genes under selection \cite{WSAT_ZivkovicandStephan2011}.
2) Are these classic expectations robust to stochastic processes affecting allele frequency dynamics at the loci under coevolution? The deterministic arms race and trench warfare dynamics can be affected by stochastic processes resulting in different polymorphism signatures than expected, and thus in a possible lower accuracy of detection. Before answering these two questions, we define the different coevolutionary models and their mathematical properties.

\subsection{The infection matrix}
An essential feature of the biology of host-parasite systems is the mechanism determining the outcome of the interaction (infection or resistance). There are $A$ alleles per species at the host and parasite interacting loci and we define $\alpha=\left( \alpha_{ij} \right)_{i,j\in\lbrace1,...,A\rbrace}$ as the infection matrix determining the compatibility of given host-parasite interactions, namely whether a host type $i$ is infected by parasite type $j$. In the literature, four models of host-parasite recognition are used (Table \ref{WSAT_infMatrices}, \textit{e.g.} \cite{WSAT_Dybdahletal2014}). The gene-for-gene (GFG) model used in plant pathology with $A=2$ supposes that one host type is susceptible to all parasite genotypes and the second one is resistant to one parasite type. In the parasite one type is recognized by the resistant host, while the other can infect both types of host. On the other hand the matching-allele (MA) model supposes a symmetric infection matrix with matching resistance/susceptibility to one parasite type. The MA model is used more often for animal-parasite interactions. We focus here in this article on the GFG model only as it shows a larger (thus more interesting) variability of coevolutionary dynamics than the MA model when changing modelling assumptions and the parameter values.

\begin{table}
	\caption{Infection matrices for four coevolution models}
		\vspace*{-12pt}\begin{center}\begin{tabular}{cccc}
			\tabularnewline\hline\tabularnewline matching-allele & inverse matching-allele & gene-for-gene & inverse gene-for-gene\\ 
			\tabularnewline\hline\tabularnewline 
			$
			\left(
			\begin{array}{rr}
			1& 0 \\
			0& 1\\
			\end{array}
			\right)
			$ 
			&
			$
			\left(
			\begin{array}{rr}
			0& 1\\
			1& 0\\
			\end{array}
			\right)
			$
			&
			$
			\left(
			\begin{array}{rr}
			0& 1\\
			1& 1\\
			\end{array}
			\right)
			$
			&
			$
			\left(
			\begin{array}{rr}
			1& 0\\
			0& 0\\
			\end{array}
			\right)
			$\\
			\tabularnewline\hline
		\end{tabular}
	\end{center}
	{The infection matrices determine the outcome of the interaction between host genotypes (rows) and parasite genotypes (columns). A successful infection is defined by $1$ and non-successful infection due to the total resistance of the host is defined as $0$. For simplicity, the rates $\alpha_{ij}$ are either chosen as one for infection or as zero for full resistance (this is an example of the $\alpha$ matrix for $A$=2).}
	\label{WSAT_infMatrices}
\end{table}
We describe in the next section how the infection matrix is integrated into two main types of mathematical models classically used: population genetics models and epidemiological models.

\subsection{Deterministic population genetics models}
Population genetic models assume a fixed population size for hosts and parasites whose genetic types change due to resampling and mutation, under a frequency-dependent disease transmission modelled as a discrete- (or continuous-) time dynamical system. For simplicity, each host individual is infected by at most one parasite type (no co-infections) \index{infectious disease model! population genetics}.
\subsubsection{Unstable model 1}
The simplest deterministic model is built assuming one parasite generation per host generation, \textit{i.e.} hosts and parasites evolve at the same time scale, (thereafter deterministic unstable model 1) yielding the following equations \index{dynamical system!discrete time}

\begin{eqnarray}
\label{WSAT_eqModel1}
{H_{i,t+1}}&=&\frac{H_{i,t}(1-c_{H_i})(1-\phi s \sum\limits_{j=1}^{A}\alpha_{ij}P_{j,t})}{\sum_{i=1}^{A}{H_{i,t}(1-c_{H_i})(1-\phi s \sum\limits_{j=1}^{A}\alpha_{ij}P_{j,t})}}=f_{i}(H_{t},P_{t}), \\
{P_{j,t+1}}&=&\frac{P_{j,t}(1-c_{P_j})\sum\limits_{i=1}^{A}\alpha_{ij}H_{i,t}}{\sum_{j=1}^{A}{P_{j,t}(1-c_{P_j})\sum\limits_{i=1}^{A}\alpha_{ij}H_{i,t}}}=g_{j}(H_{t},P_{t}). \nonumber
\end{eqnarray}

We define $H_{t}=\left(H_{i,t}\right)_{i\in \lbrace1,...,A\rbrace}$ as the frequency of host types $i$ in the host population at generation $t$ and $P_{t}=\left(P_{j,t}\right)_{j\in \lbrace 1,...,A\rbrace}$ for parasite type $j$ in the parasite population. Thus ${H_{i,t+1}}$ and ${P_{j,t+1}}$ are the frequencies in the next generation $t+1$. Fitness costs associated with given alleles can be defined for the host of type $i$ ($c_{H_i}$) and the parasite type $j$ ($c_{P_j}$) due to trade-offs between resistance/infectivity and fitness. The fitness cost of any host infected by any parasite is denoted as $s$, and $\phi$ is the proportion of hosts receiving parasites. \\
\indent In their common form, the coevolutionary models assume thus that one or few major genes determine the outcome of interactions \cite{WSAT_MayandAnderson1983, WSAT_TellierandBrown2007a, WSAT_Dybdahletal2014}. These models are well suited for 1) host species with discrete generations such as annual plants, 2) parasites which undergo few generations per host generation, and 3) parasites whose infection rate and infectious period depend strongly on the environment, as is common for plants or invertebrates. Indeed, in the latter case, $\phi$ can be variable depending on environmental conditions such as temperature or humidity, and only weakly dependent on the amount of infected hosts at the previous generation (the disease prevalence). The system of equations~\eqref{WSAT_eqModel1} can be written for a bi-allelic GFG system ($A=2$) as shown in Table \ref{WSAT_infMatrices} for host type 1 (so-called resistance allele) and parasite type 2 (so-called infectivity allele) while assuming $0 < c_{H_1}, c_{P_2} <1$ and $c_{H_2}=c_{P_1}=0$:
\begin{eqnarray}
\label{WSAT_eqGFGModel1}
{H_{1,t+1}}&=&\frac{H_{1,t}(1-c_{H_1})(1-\phi s P_{2,t})}{H_{1,t}(1-c_{H_1})(1-\phi s P_{2,t})+H_{2,t}(1-\phi s)}, \\
{P_{2,t+1}}&=&\frac{P_{2,t}(1-c_{P_2})}{P_{1,t}H_{2,t}+P_{2,t}(1-c_{P_2})}. \nonumber
\end{eqnarray}
We can compute the following polymorphic equilibrium point frequencies for the GFG deterministic model 1 at which ${H_{i,t+1}}=H_{i,t}=\widehat{H_{i}}$ and ${P_{j,t+1}}=P_{j,t}=\widehat{P_{j}}$:
\begin{eqnarray}
\label{WSAT_eqpointModel1}
\widehat{H}_{1}=c_{P_2} \text{ and } \widehat{P}_{2}=\frac{\phi s -c_{H_1}}{\phi s(1-c_{H_1})}.
\end{eqnarray}
	
The deterministic model 1 (equation~\eqref{WSAT_eqModel1}) is driven only by niFDS and the trace of its Jacobian matrix evaluated at the polymorphic equilibrium is zero and the eigenvalues are complex numbers with zero real parts \cite{WSAT_TellierandBrown2007a}. For the case $A=2$, the internal polymorphic equilibrium (equation~\eqref{WSAT_eqpointModel1}) is indeed an unstable saddle point \cite{WSAT_Kot2001} and over time one observes cycling of allele frequencies moving away from this point. This also means that one of the four monomorphic equilibria ((0,0), (1,0), (0,1) or (1,1)) is the stable attractor \cite{WSAT_TellierandBrown2007a}. The behaviour of the dynamical system is then referred to as being unstable characterized by only transient presence of several types at the host and parasite coevolving loci.

\subsubsection{Stable model 2}
Additional hypotheses which change the stability property of the internal equilibrium can be added to equation \eqref{WSAT_eqModel1} (reviewed in \cite{WSAT_BrownTellier2011}). Several life-history traits of the host or the parasite can generate a stable polymorphic attractor such as the host exhibiting seed banks (a delay in seed germination by two or more years, \cite{WSAT_TellierandBrown2009, WSAT_verin2018}), or the parasite undergoing several generations per host generation with auto-infection (parasites infecting the same host, \cite{WSAT_TellierandBrown2007a}). We explicitly describe here an extension of the bi-allelic GFG model 1 with two parasite generations per host generation with only auto-infection and 100\% of the hosts receiving parasites ($\phi=1$). Auto-infection is defined as the probability for a parasite offspring to remain on the same host individual (here one). In this model, the parasite evolves on a faster time-scale than the host. The other parameters are defined as for model 1. We define $s_1$ as the loss of fitness of plants infected during one parasite generation, and $s_2$ when infected by two consecutive parasite generations (within a host generation $t$). Two equations are needed for the parasite population to describe the frequency of parasite type $j$ after the first parasite generation ($\breve{P_{j,t}}$) and at the next host generation (${P_{j,t+1}}$) \index{dynamical system!discrete time}

\begin{equation}
\label{WSAT_eqGFGModel2}
\begin{aligned}
\breve{P_{2,t}} &= \frac{P_{2,t}(1-c_{P_2})} {P_{2,t}(1-c_{P_2}) + P_{1,t} H_{2,t}},\\
{H_{1,t+1}} &= \frac{H_{1,t} (1-c_{H_1}) [P_{1,t} \breve{P_{1,t}} +  P_{1,t} \breve{P_{2,t}} (1-s_2)  + P_{2,t} (1-s_1)]} {H_{1,t} (1-c_{H_1}) [P_{1,t} \breve{P_{1,t}} +  P_{1,t} \breve{P_{2,t}} (1-s_2) + P_{2,t}(1-s_1) +H_{2,t}(1-s_1)]},\\
{P_{2,t+1}} &= \frac{(1-c_{P_2}) [ H_{1,t} (P_{1,t} \breve{P_{2,t}} + P_{2,t}) + H_{2,t} P_{2,t}]} {(1-c_{P_2}) [H_{1,t} (P_{1,t} \breve{P_{2,t}} + P_{2,t})  + H_{2,t} P_{2,t}]  + H_{2,t} P_{1,t}}.
\end{aligned}
\end{equation}
The equilibrium frequencies $\widehat{H_{1}},\,\widehat{P_{2}}$ at the internal polymorphic equilibrium point are computed assuming $c_{H_1}^{2}<<c_{H_1}$ and $c_{P_2}^{2}<<c_{P_2}$ so that $c_{H_1}^{2}$ and $c_{P_2}^{2}$ can be neglected \cite{WSAT_TellierandBrown2007a}:
\begin{equation}
\label{WSAT_eqpointModel2}
\begin{aligned}
\widehat{H_{1}} &\approx \frac{c_{P_2}}{2-c_{P_2}-\widehat{P_{2}}},\\
\widehat{P_{2}} &\approx \frac{s_2 + s_1 -\sqrt{(s_2 + s_1)^2 - 4s_2(s_1-c_{H_1})}}{2s_2 (1-c_{H_1})}.\\
\end{aligned}
\end{equation}

Model 2 features so-called negative direct FDS (ndFDS), the trace of the Jacobian evaluated at the equilibrium point becomes negative and the eigenvalues of the Jacobian matrix are complex numbers with negative real parts \cite{WSAT_TellierandBrown2007a, WSAT_Kot2001} \index{dynamical system!Jacobian matrix}. In discrete-time, the polymorphic equilibrium point is a stable attractor if the eigenvalues lie within a unit circle centred on (-1,0) in the complex plane, which can occur for some parameter combinations (example in Figure \ref{WSAT_Fig1}a, \cite{WSAT_TellierandBrown2007a}). In this case, long-term stable polymorphism defined as the maintenance of several types over time is promoted in host and parasite populations at the coevolving loci. Over time, one observes  cycling of allele frequencies damping off towards the internal equilibrium point. \\
Seed banking may generate ndFDS as well, and we have further shown that seed banks can evolve as a bet hedging strategy as a response to host-parasite coevolutionary dynamics \cite{WSAT_verin2018}. It is thus of interest to predict the consequences of seed banking for host and/or parasite nucleotide diversity and for allele frequencies at the coevolving loci. The model of seed bank in \cite{WSAT_TellierandBrown2009} assumes unbounded geometric time for seeds to be dormant which according to the work of  \cite{WSAT_blath2015} should generate multiple merger coalescent processes. In contrast, when a bounded time is assumed for the maximum time of seed dormancy \cite{WSAT_verin2018}, a rescaled Kingman $n$-coalescent process is expected for the host population \cite{WSAT_kajetal2001} whose parameters can be estimated using full genome data \cite{WSAT_sellinger2020}. \index{coalescent!seed banks} ndFDS is obtained in \cite{WSAT_verin2018} by assuming non-geometric shapes of the seed dormancy function. More details on the coalescent processes under seed banks are found in \cite{WSAT_blathKurt2019, WSAT_GrevenAnddenHollander2019}.

\subsection{Deterministic epidemiological model}
More complex epidemiology models are also used \cite{WSAT_KermackandMcKendrick1927, WSAT_MayandAnderson1983}, in which host and parasite population sizes are variable in time and disease transmission is density-dependent. In analogy to \cite{WSAT_Dybdahletal2014}, the changes in numbers (or masses) of host and parasite individuals over time are determined by the following coupled differential equations (hereafter model 3) \index{dynamical system!continuous time}:
\begin{equation}
\label{WSAT_eq3}
\begin{aligned}
\frac{dH_i}{dt}&=H_i \left[b_i(1-c_{H_i})-d_i-\sum\limits_{j=1}^{A}\alpha_{ij}\beta_{ij}(1-c_{P_j})\sum\limits_{k=1}^{A}I_{kj}\right]\\
&+b_i(1-c_{H_i})(1-s)\sum\limits_{j=1}^{A}I_{ij},\\
\frac{dI_{ij}}{dt}&=I_{ij}(-d_i-\delta)+H_i\left[\alpha_{ij}\beta_{ij}(1-c_{P_j})\sum\limits_{k=1}^{A}I_{kj}\right].
\end{aligned}
\end{equation}

\noindent{} In Equation \eqref{WSAT_eq3}, $H_i$ are the number of healthy (\textit{i.e.}, non-infected) individuals of genotype $i$, and $I_{ij}$ denote the number of hosts of genotype $i$ infected by a parasite of genotype $j$. $b_i$ and $d_i$ are the birth and natural death rates (\textit{i.e.}, independent of the disease) of host genotype $i$, respectively, and $\delta$ is the disease-induced death rate caused by pathogens on infected hosts (\textit{i.e.} the effect of pathogen on host mortality). $\beta_{ij}$ is the disease transmission rate between a parasite of genotype $j$ and a host of genotype $i$. As for model 1, we define $c_{H_i}$ and $c_{P_j}$ as the costs for the hosts and the parasites of carrying genotype $i$ and $j$, respectively, and $s$ is the decrease of host reproductive fitness due to an infection (\textit{i.e.} the effect of the pathogen on host fecundity). \index{infectious disease model! epidemiology}
The assumptions are here that parasites only survive inside infected hosts, and that there is 1) overlap of host generations and 2) overlap between infectious and healthy hosts for disease to transmit. The disease prevalence is here defined by the dynamics of the system and does vary in time depending on the host and parasite allele frequencies. A discrete time equivalent can also be obtained by analogy to Lotka-Volterra types of equations \cite{WSAT_Gokhaleetal2013}. These epidemiological models are characterized by ecological-evolutionary feedbacks which generates ndFDS, and thus often exhibit a stable internal equilibrium point or stable cycling of allele frequencies \cite{WSAT_AshbyBoots2017,WSAT_zivkovic2019}. The monomorphic and polymorphic equilibrium points for the deterministic model 2 can be computed and are found in \cite{WSAT_zivkovic2019}, though we refrain to show them here as these are cumbersome and less intuitive than those of model 1.

\subsection{Aims of the present study}
\indent Coevolutionary models have thus been concerned with predicting and assessing the stability of polymorphism at the coevolving loci based on these various recognition models. Generally, GFG shows more unstable allele dynamics than MA models. However, the role and influence of stochasticity on the model behaviour has been fairly neglected. Indeed, stochasticity will generate departures for GFG and MA from the deterministic outcomes by affecting the allele frequency trajectories and thus the maintenance of alleles in both populations (fixation or loss) as well as the period and amplitude of coevolutionary cycles. As our aim is to eventually connect DNA polymorphism sequences that can be obtained from hosts and parasites, we have investigated 1) if and how stochasticity modifies predictions of the deterministic models, and 2) what the expected polymorphism signatures of coevolution are at the loci under coevolution and over the whole host and parasite genomes. In the following we describe all stochastic processes occurring in host-parasite coevolution and their influence on the system behaviour. We distinguish two types of stochastic processes. Intrinsic processes originate from the population model themselves and are constituted by 1) genetic drift and mutation in host and parasite finite populations, and 2) stochastic disease transmission among individuals. Extrinsic stochastic processes are generated by spatial structuring of populations and gene flow, environmental stochasticity and higher-order interactions of host and parasite species within ecological networks. As extrinsic stochastic factors were outside the scope of our project, we propose here only a short summary of open questions as their study has been largely neglected in the literature.

\section{Intrinsic stochasticity}
In the following we introduce two stochastic processes, namely genetic drift and mutation, into the deterministic population genetics and epidemiological models described above. 

\subsection{Genetic drift and mutation at the coevolutionary loci}
\subsubsection{Definitions and modelling framework}
\indent In \cite{WSAT_Tellieretal2014}, random stochastic reproduction as modelled by a classic Wright-Fisher (WF) population process affects the deterministic predictions of model 1 and model 2. To reveal the signatures of coevolution at the interacting loci, we model two types of stochastic processes: 1) the drift affecting allele frequency change, and 2) the coalescent process within each type. We thus couple a backward in time Kingman $n$-coalescent simulator to the forward coevolution simulations with genetic drift (we describe below the results originally published in \cite{WSAT_Tellieretal2014}). We first include the role of genetic drift which nudges the coevolutionary dynamics, \textit{i.e.} allele frequency trajectories, away from the deterministic one from model 1 or 2. Host and parasite populations have fixed size $N_{H}$ and $N_{P}$, respectively. \index{genetic drift! Wright-Fisher model}
The forward simulations under a bi-allelic model ($A=2$) follow a WF model assuming binomial sampling at each generation in the host and parasite population based on the deterministic value of the types' frequencies ($f_1, g_1$ defined in equation ~\ref{WSAT_eqModel1}) given in model 1 (equation ~\ref{WSAT_eqGFGModel1}) or model 2 (equation ~\ref{WSAT_eqGFGModel2}). The initial types' frequencies are drawn randomly in $]0,1[$. The frequency of host type $i$ ($\cH_t=\left( \cH_{i,t} \right)_{i\in \lbrace 1,2 \rbrace}$) and parasite type $j$ ($\cP_t=\left( \cP_{j,t} \right)_{j\in \lbrace 1,2 \rbrace}$) under the model 1 (eq.~\ref{WSAT_eqModel1}) with genetic drift are given at $t+1$ by:
\begin{equation}
\label{WSAT_eqBinomial}
\begin{aligned}
\cH_{1,t+1}\sim B (N_H,f_1(\cH_t,\cP_t))/N_H, \\
\cP_{1,t+1}\sim B (N_P,g_1(\cH_t,\cP_t))/N_P.
\end{aligned}
\end{equation}

We generate an example of such forward in time dynamics under model 2 in Figure \ref{WSAT_Fig1}b.\\

\indent We also assume that recurrent mutations can occur between the two types (as $A=2$) in hosts and parasites \cite{WSAT_Tellieretal2014}. The mutational step is thus a binomial sampling for each individual ($N_H$ hosts and $N_P$ parasites) which is approximated by a Poisson process defined by the probabilities to mutate between the two alleles. The mutation probabilities are $\mu_{H,k \rightarrow i}$ to mutate from type $k$ to $i$ in the host (the reverse mutation has probability $\mu_{H,i \rightarrow k}$), and $\mu_{P,l \rightarrow j}$ to mutate from $l$ to $j$ in the parasite (and $\mu_{P,j \rightarrow l}$ for the reverse). The number of hosts of type $i$ at the next generation is simply the sum of the $i$ individuals which did not mutate and of the type $k$ individuals which did mutate (and similarly in the parasite). \\
This forward in discrete time simulation of type frequency dynamics constrains a backward in time coalescent process applied to the host and parasite populations with two types. The produced genealogy of the discrete time Wright-Fisher dynamics is thus approximated by the continuous time Kingman $n$-coalescent. \index{coalescent! Kingman} The sample size, \textit{i.e.} the number of sequenced individuals, of hosts and parasites are at present, respectively, $n_{H}$ and $n_{P}$. This stochastic model produces an approximated structured coalescent \cite{WSAT_Nordborg1997} for the host and for the parasite population, each with two types whose frequencies vary over time and with recurrent mutation at the coevolutionary locus (here mutation between types corresponds to migration between demes in the spatially structured coalescent).\\

\indent Finally, following \cite{WSAT_Tellieretal2014} we distribute on the coalescent tree without recombination neutral segregating mutations (around the site under coevolution) under an infinitely-many mutation model following the classic Poisson process with locus rate $\Theta_{H}=4N_{H} \mu_{H_{neutral}}$ and $\Theta_{P}=4N_{P} \mu_{P_{neutral}}$ for hosts and parasites, respectively. The polymorphism footprints of coevolution are assessed by computing several statistics for host and parasite based on all selected and neutral segregating sites (SNPs). The unfolded site-frequency spectrum (SFS) is one of the most commonly used statistics for the analysis of segregating SNPs. It is defined as the distribution of the number of times $u$ a mutation is observed in a sample of $n$ sequences and the site frequencies are denoted as $f_{n,u}$ ($1 \leqslant u \leqslant n-1)$. Two related measures can be computed from the neutral SNPs: the expected number of segregating sites $S_{n}$ which equals the total number of mutations in the infinitely-many sites model, and the average number of pairwise differences $\Pi_{n}$.
\begin{equation}
S_{n}= \sum_{u=1}^{n-1} f_{n,u} \text{ , and }  \Pi_{n}=\frac{1}{\binom{n}{2}} \sum_{u=1}^{n-1}u(n-u)f_{n,u}.\\ \nonumber 
\end{equation}

In the present paper, we introduce another drift model based on the $\Psi$-coalescent model \cite{WSAT_eldon2006} \index{coalescent! $\Psi$-coalescent}. Genetic drift is generated as a discrete-time Moran model in the host and parasite population by replacing one dying individual at each time-step by the offspring of any of the parents (including itself), until a large reproductive events occurs \index{genetic drift! Cannings model}. At this event, a given proportion $\Psi_{H}$ of host or $\Psi_{P}$ of parasite individuals are replaced by the offspring from one randomly chosen parent of a given type (\cite{WSAT_eldon2006}, a special case of a Cannings model \cite{WSAT_cannings1974}). This step yields values of $\cH_{1,t+1}$ and $\cP_{1,t+1}$ under the $\Psi$-coalescent model (example of dynamics under model 2 in Figure \ref{WSAT_Fig1}c,d). \\

\subsubsection{Results}
\begin{figure}[b!]
	(a) \subfloat{\includegraphics*[width=0.43\textwidth]{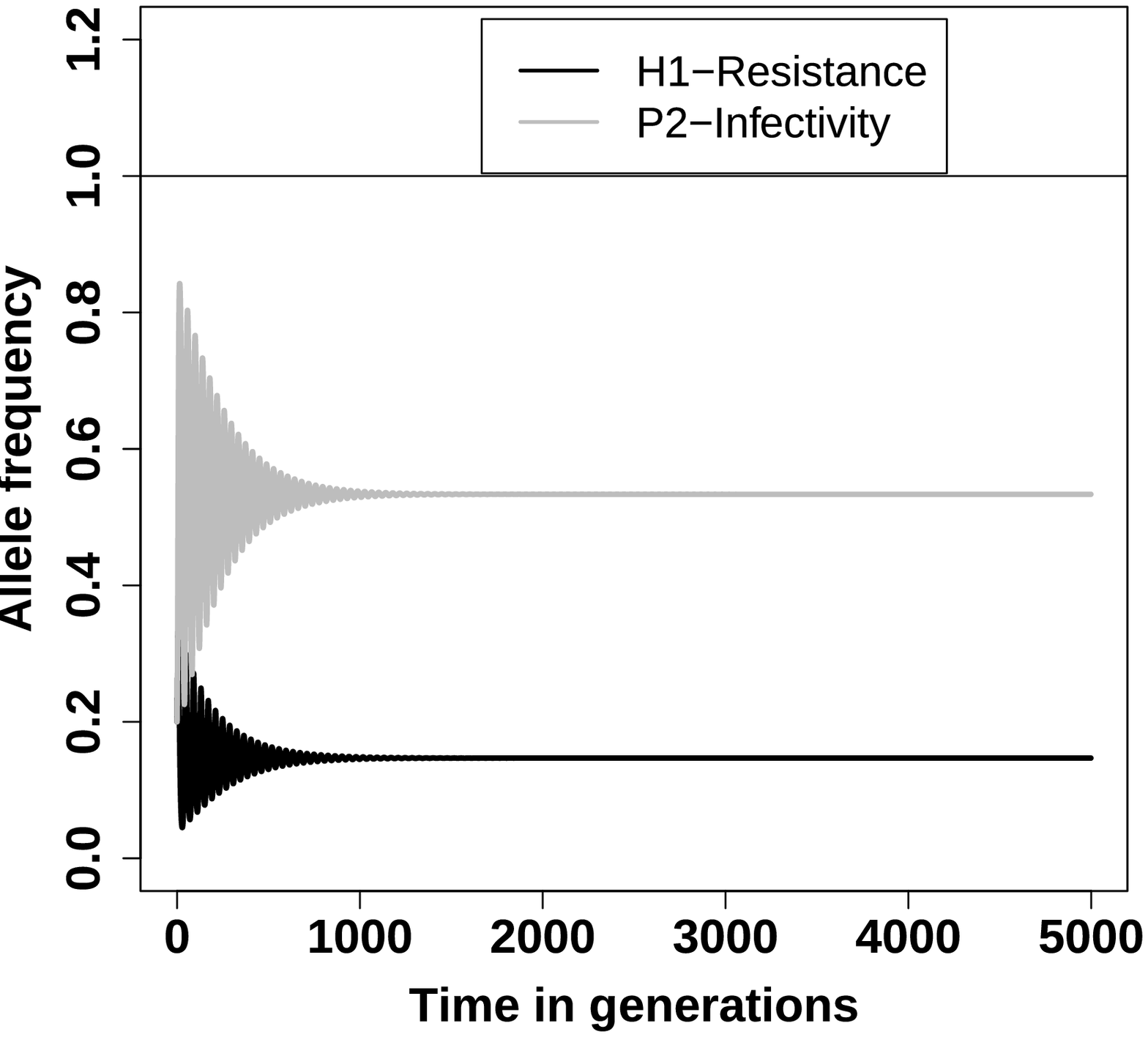}}\quad
	(b) \subfloat{\includegraphics*[width=0.43\textwidth]{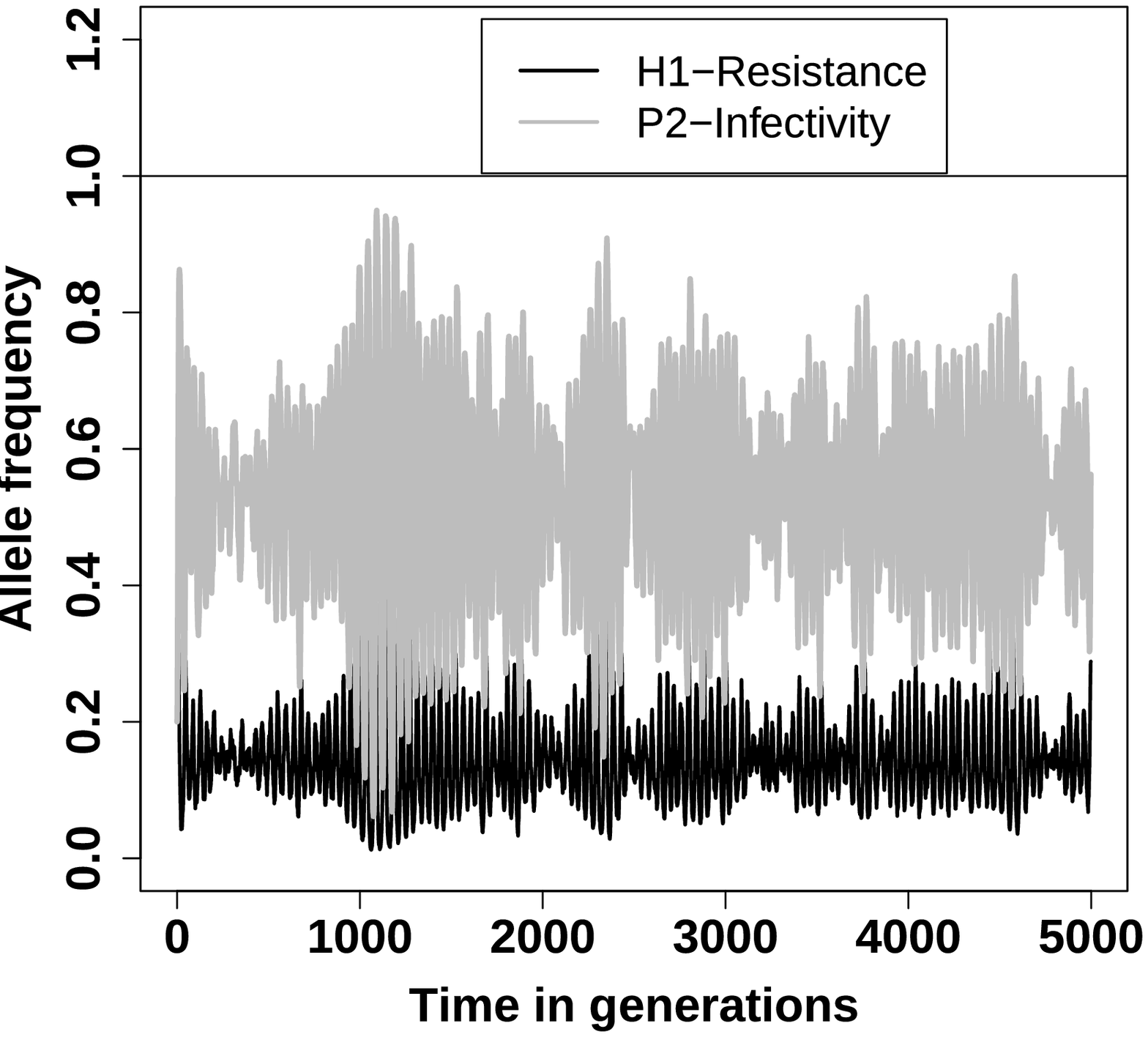}}\quad
	(c) \subfloat{\includegraphics*[width=0.43\textwidth]{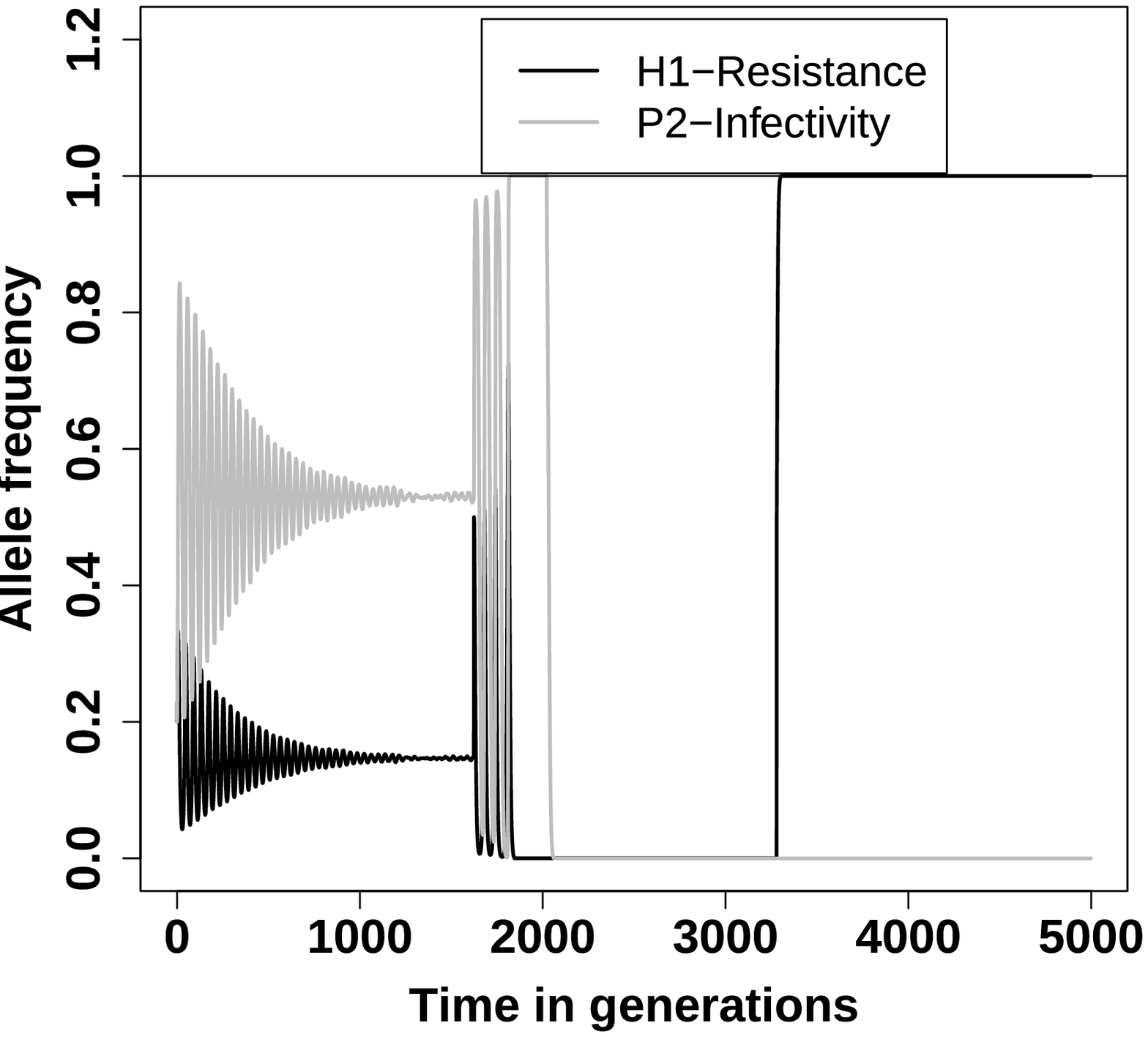}}\quad
	(d) \subfloat{\includegraphics*[width=0.43\textwidth]{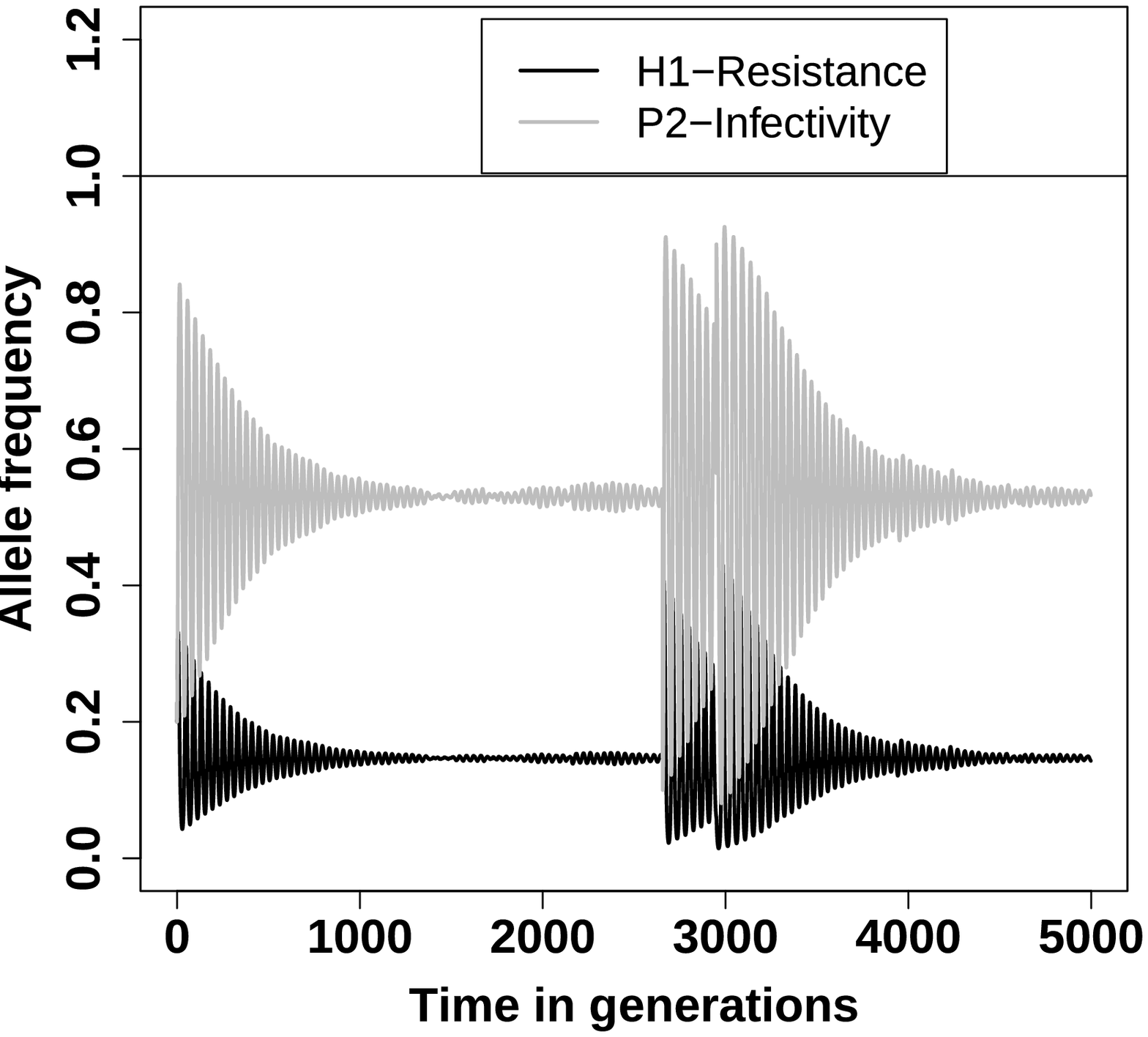}}\quad
	\caption{Parasite infectivity ($P_{2}$) and host resistance ($H_{1}$) alleles are plotted for the GFG model of Table \ref{WSAT_infMatrices} over time by numerically solving the equations \eqref{WSAT_eqGFGModel2} for the following parameters $c_{H_1}=0.2$, $c_{H_2}=0$ $c_{P_1}=0$, $c_{P_2}=0.2$, $s=0.4$, and the population sizes $N_{H}=N_{P}=3000$. The stochastic simulations are obtained with $\mu_{H,1 \rightarrow 2}$=$\mu_{H,2 \rightarrow 1}$=$\mu_{P,1 \rightarrow 2}$=$\mu_{P,2 \rightarrow 1}$=$10^{-5}$ and $\mu_{H,neutral}=\mu_{P,neutral}=25\times10^{-6}$. (a - top left) The deterministic allele frequencies converge to the stable internal polymorphic equilibrium point. (b - top right) With genetic drift implemented as a Wright-Fisher model, stochastic oscillations do occur around the equilibrium frequencies. (c - bottom left) and (d - bottom right) Two simulation outcomes are shown under the $\Psi$-coalescent model with the probability of large reproductive events to occur being $1/N_{H}$ and $1/N_{P}$ in hosts and parasites, respectively. The proportion of individuals replaced by the offspring from one randomly chosen parent during these events is in hosts $\Psi_{H}=0.5$ and in parasites $\Psi_{P}=0.9$.}
	\label{WSAT_Fig1}
\end{figure}

\indent Model 1 always yields arms race dynamics characterized by fixation of alleles in the host and parasite population at the coevolving loci \cite{WSAT_Woolhouseetal2002, WSAT_Holub2001}. Such arms race dynamics are due to recurrent selective sweeps \index{selection! positive} occurring at the loci because 1) types reach fixation due to genetic drift, and 2) we have introduced recurrent forward-backward mutation between types. In model 2 with an internal stable polymorphic equilibrium, the so-called trench warfare dynamics is expected to be seen with the maintenance of several alleles in the host and parasite populations over long periods of time. \index{selection! balancing} In this case it is expected that signatures of balancing selection are observed at the coevolving loci \cite{WSAT_Woolhouseetal2002, WSAT_Holub2001}.
The first result we obtained \cite{WSAT_Tellieretal2014} is that the variance in the observed polymorphism signatures at the coevolutionary loci between simulations under the same coevolutionary parameters (as measured by the SFS, $S_{n_H}$, $S_{n_P}$, $\Pi_{n_H}$, and $\Pi_{n_P}$) is largely affected by genetic drift changing the allele frequency dynamics rather than by the coalescent process. For example, balancing selection was observed only for a small range of the parameter space under model 2 (under the WF drift model). Indeed, even though the internal equilibrium is for some parameter combinations deterministically stable, the populations never reach it because genetic drift leads to fixation of alleles in the host and in the parasite population (model 2). Indeed, under small population sizes ($N_{H}$ or $N_{P}$ $<$ 2000) genetic drift counter-acts balancing selection resulting in the fixation of one type \cite{WSAT_Tellieretal2014}. \index{genetic drift! Wright-Fisher model}
A new result we show here, is that this effect of genetic drift is less prominent if only a fraction of the population is replaced at each generation, \textit{i.e.} assuming overlap of host or parasite generations. Allele frequency cycles with small amplitude around the equilibrium point are observed between large reproductive events in model 2 with $\Psi$-coalescent (Fig. \ref{WSAT_Fig1}~c,d, and also \cite{WSAT_TellierandBrown2007a}). A similar effect of genetic drift on allele frequency trajectories and allele fixation was observed in \cite{WSAT_Gokhaleetal2013} in a simpler version of our model 3. When including possible large reproductive events which affect a large proportion of the offspring as modelled in the $\Psi$-coalescent, \index{coalescent! $\Psi$-coalescent} the number of possible types of model behaviour increases, ranging from fixation of alleles to damping off cycle dynamics sustained by genetic drift (Fig. \ref{WSAT_Fig1}~c,d) \index{genetic drift! Cannings model}. \\

\indent A second result from \cite{WSAT_Tellieretal2014} is that under the GFG model of equation \eqref{WSAT_eqGFGModel2}, even if balancing selection occurs, the equilibrium frequency of one host allele exhibits a low frequency (here the host resistant allele 1). This effect would not be occurring under the MA model where both host types have similar frequencies at equilibrium (because $c_{P_1}=c_{P_2}$). As a result, under GFG, the host resistant allele has a very small population size and the underlying coalescent genealogy is imbalanced and typical signatures of balancing selection are not observed. We thus predict that under the GFG model, balancing selection will be very difficult to observe at the host coevolving locus, and should be preferably studied at the parasite locus. \index{selection! balancing} In general, polymorphism signatures are stronger in parasite than in host samples \cite{WSAT_Tellieretal2014}.

We then investigated in \cite{WSAT_Tellieretal2014} the period and amplitude of cycles for a large range of parameter combinations. Under arms race dynamics, the period of cycles is determined by the population mutation rate of host and parasite at the coevolving locus because after a selective sweep the waiting time for the introduction of a new allele depends on $4N_{H}\mu_{H,i \rightarrow k}$ and $4N_{H}\mu_{H,k \rightarrow i}$ in the host population (respectively $4N_{P}\mu_{P,j \rightarrow l}$ and $4N_{P}\mu_{P,l \rightarrow j}$ in the parasite population). In the trench warfare scenario, the period and amplitude of cycles are small only for a reduced range of the parameter space, in which polymorphism is maintained in both antagonistic species at intermediate frequencies. In contrast to a previous suggestion \cite{WSAT_Woolhouseetal2002}, we conclude that fast coevolutionary cycles with small amplitude are only observed over a small range of parameter values and thus cannot be considered as hallmarks of trench warfare dynamics \cite{WSAT_Tellieretal2014}.
\\ \indent We have recently extended these results showing by simulations that unique combinations of polymorphism signatures are generated under arms race or trench warfare for different parameter combinations. It is thus possible to infer the model parameters based on polymorphism data at the coevolving loci if the population sizes of hosts and parasites are known \cite{WSAT_maerkle2019}. Very good accuracy of estimation of the parameters $s$, $c_{H}$ and $c_{P}$ was obtained with an approximate Bayesian computation method using jointly host and parasite polymorphism data and from at least 10 repetitions of the coevolution run, \textit{i.e.} for laboratory controlled experiments \cite{WSAT_maerkle2019}.

\subsection{Genetic drift and mutation at multiple coevolutionary loci}
Although not part of our investigation, it is of interest to review here the effect of genetic drift and mutation in multi-locus coevolving systems. This situation is indeed realistic as many hosts are known to harbour several hundreds of resistance genes (nod-like receptors in plants or invertebrates). However, it is yet unclear how many of these genes are key for resistance to a given pathogen species, as it often appears that only a handful have major effects on the resistance phenotype. Models 1 and 2 can be extended to more host and parasite genotypes either by including several loci (in linkage equilibrium or disequilibrium, \cite{WSAT_TellierandBrown2007b, WSAT_salathe2005}) or several alleles per locus.
The expected behaviour of the dynamical system of model 1 in this case does not change with the introduction of more loci \cite{WSAT_TellierandBrown2007b}, and one genotype of host and parasite ultimately gets fixed due to genetic drift (arms race). However, if mutation rate is high enough, polymorphism can be observed in one interacting species while the other species is monomorphic \cite{WSAT_salathe2005}. This occurs for example in a GFG model as a super-virulent parasite, which can infect all hosts, is fixed while in the host, polymorphism for different multi-locus genotypes occurs because these have identical fitness (and a high mutation rate). In fact, under a multi-locus coevolution model, genetic drift and mutation can counter-act the action of selection, which would lead to allele fixation, and so-called statistical polymorphism can occur \cite{WSAT_allen1975}. However, we expect that this statistical polymorphism is a quasi-state which is unstable and transient, resulting ultimately in signatures of arms race (selective sweeps) at the polymorphism level and not of trench warfare (balancing selection). In model 3 with many loci, a stable polymorphic equilibrium arises \cite{WSAT_MayandAnderson1983, WSAT_AshbyBoots2017}, and we speculate that a mutation-drift-coevolution equilibrium may exist where many genotypes are maintained around the equilibrium frequencies by the action of drift and mutation introducing new variants.

\subsection{Genetic drift and mutation over the genomes of interacting species}
\subsubsection{Definitions and modelling framework}
\indent So far we have studied the polymorphism signatures of coevolution at loci underlying coevolution. This is legitimate as the aim of empirical population genetics is to detect such loci in the genomes of host and parasites. However, in some models (model 3) host and parasite population size may vary in time due to the epidemiological feedbacks \cite{WSAT_AshbyBoots2017}. Such changes in population sizes are termed as the co-demographic history of hosts and parasites \cite{WSAT_zivkovic2019}. \index{demography! host-parasite co-demography} Our aim was to study the genome-wide signatures (SFS, $\Pi$) in the host and parasite populations. In other words and in contrast to classic population genetics, we investigate the neutral patterns of diversity in host and parasite genomes due to demographic changes generated by coevolution occurring at one locus.
We use a theoretical framework based on a forward diffusion process in a neutral Wright-Fisher model with stepwise change in population size as developed in \v{Z}ivkovi{\'c} and Stephan \cite{WSAT_ZivkovicandStephan2011}. \index{genetic drift! Wright-Fisher model} This allows us to derive  the site-frequency spectrum (SFS) of neutral alleles. We obtain semi-analytical results for the SFS over time using Mathematica \cite{WSAT_zivkovic2019}. We discretise the continuous time dynamics of model 3 into very small time steps at which we can compute the change in SFS. We investigated first the behaviour of the dynamical system of model 3, and choose the parameter space for which cycles of coevolution occur, \textit{i.e.} cycling of allele frequencies and of host and parasite population changes. Note that a large number of parameter combinations yield either monomorphic populations or a stable polymorphic equilibrium at which population sizes are constant. These cases are not of interest here as the outcome collapses to model 1. \\

\subsubsection{Results}
\begin{figure}[b!]
	(a) \subfloat{\includegraphics*[width=0.43\textwidth]{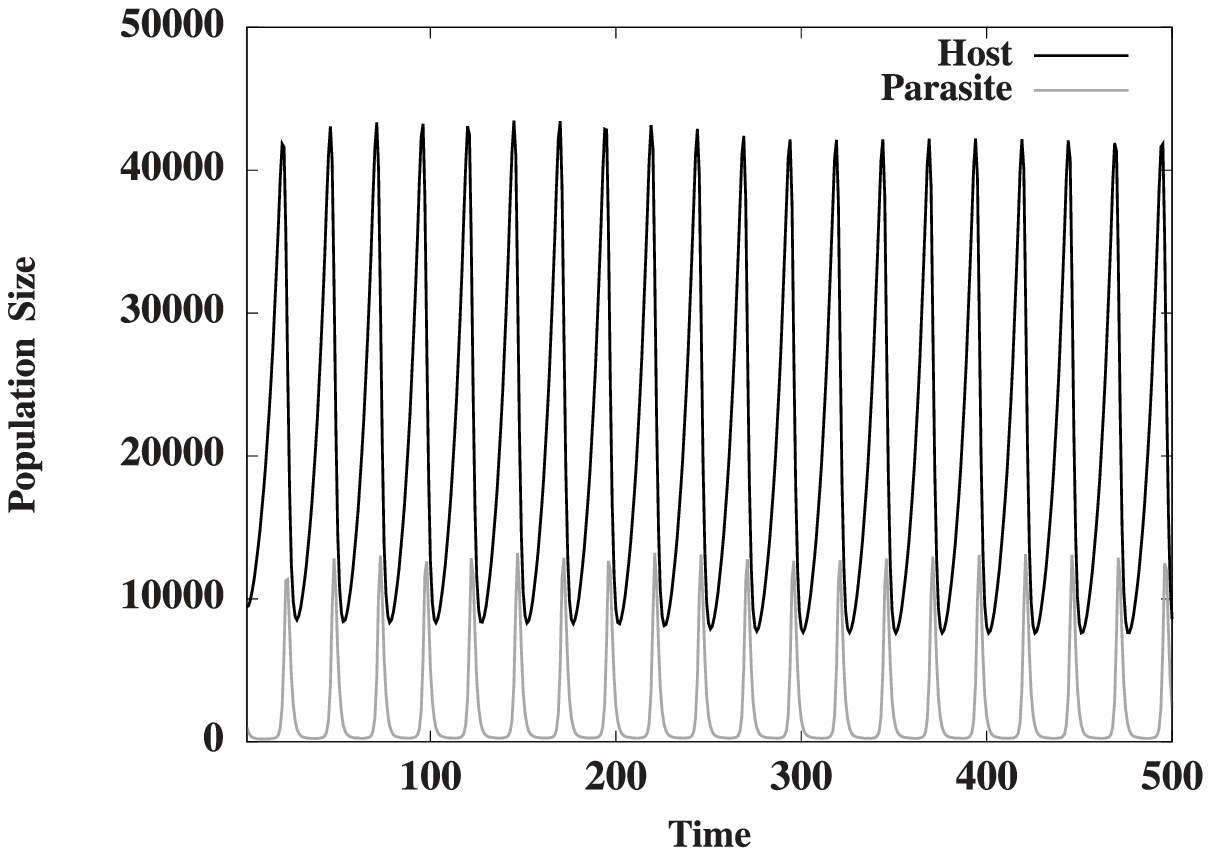}}\quad
	(b) \subfloat{\includegraphics*[width=0.43\textwidth]{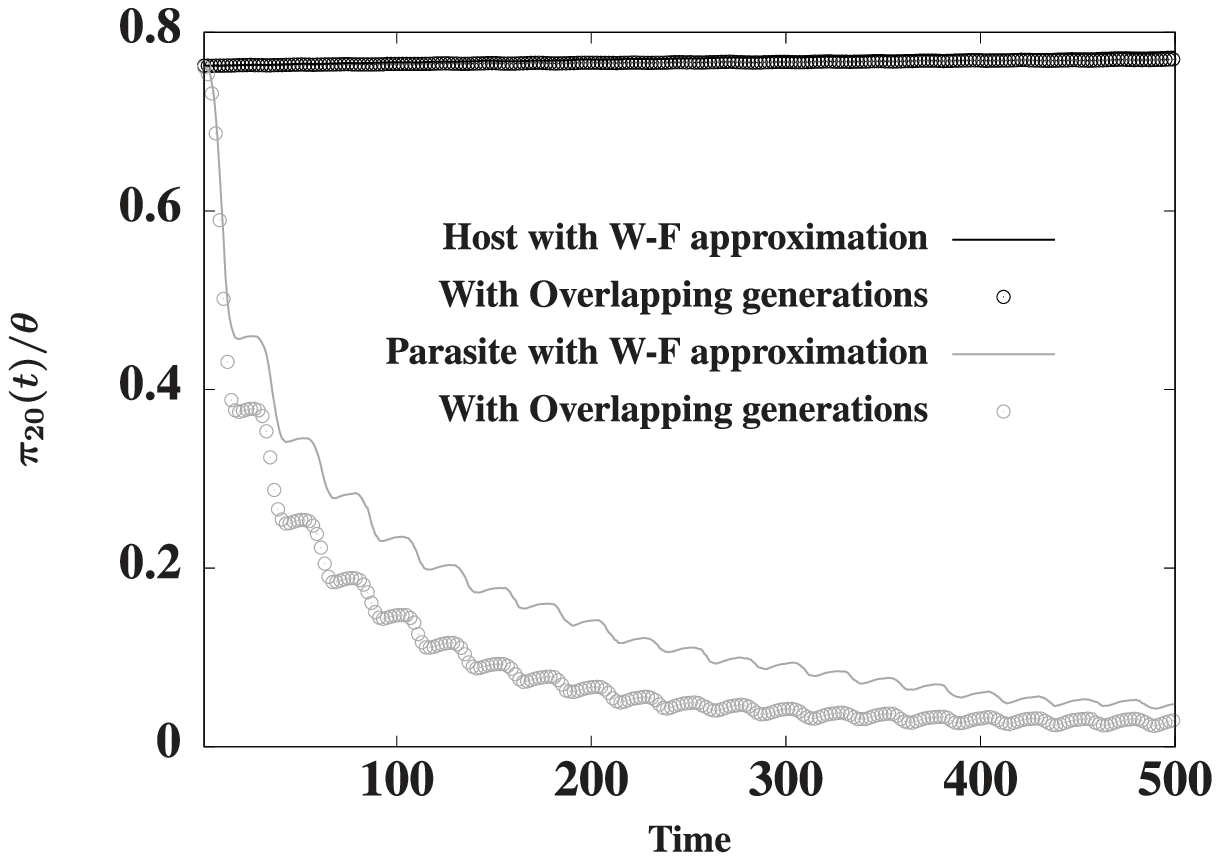}}\quad
	(c) \subfloat{\includegraphics*[width=0.43\textwidth]{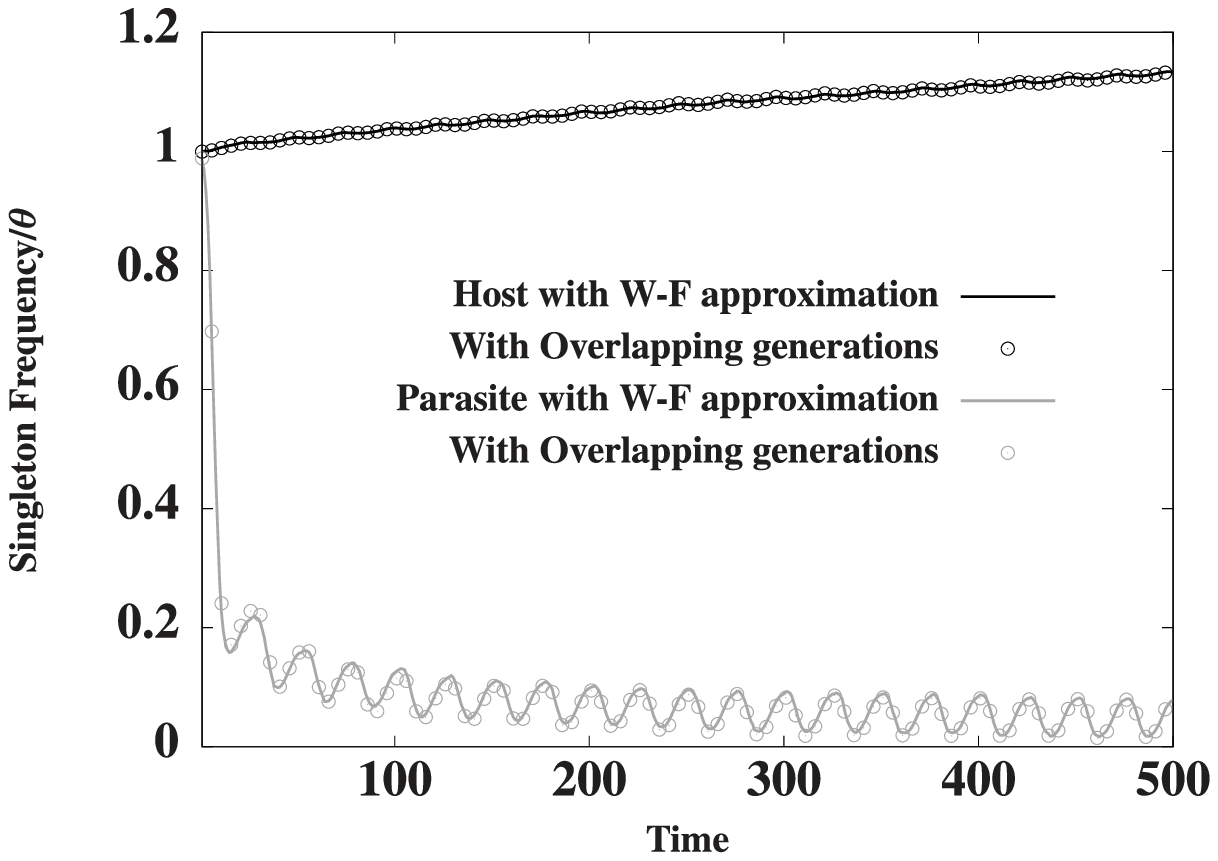}}\quad
	(d) \subfloat{\includegraphics*[width=0.43\textwidth]{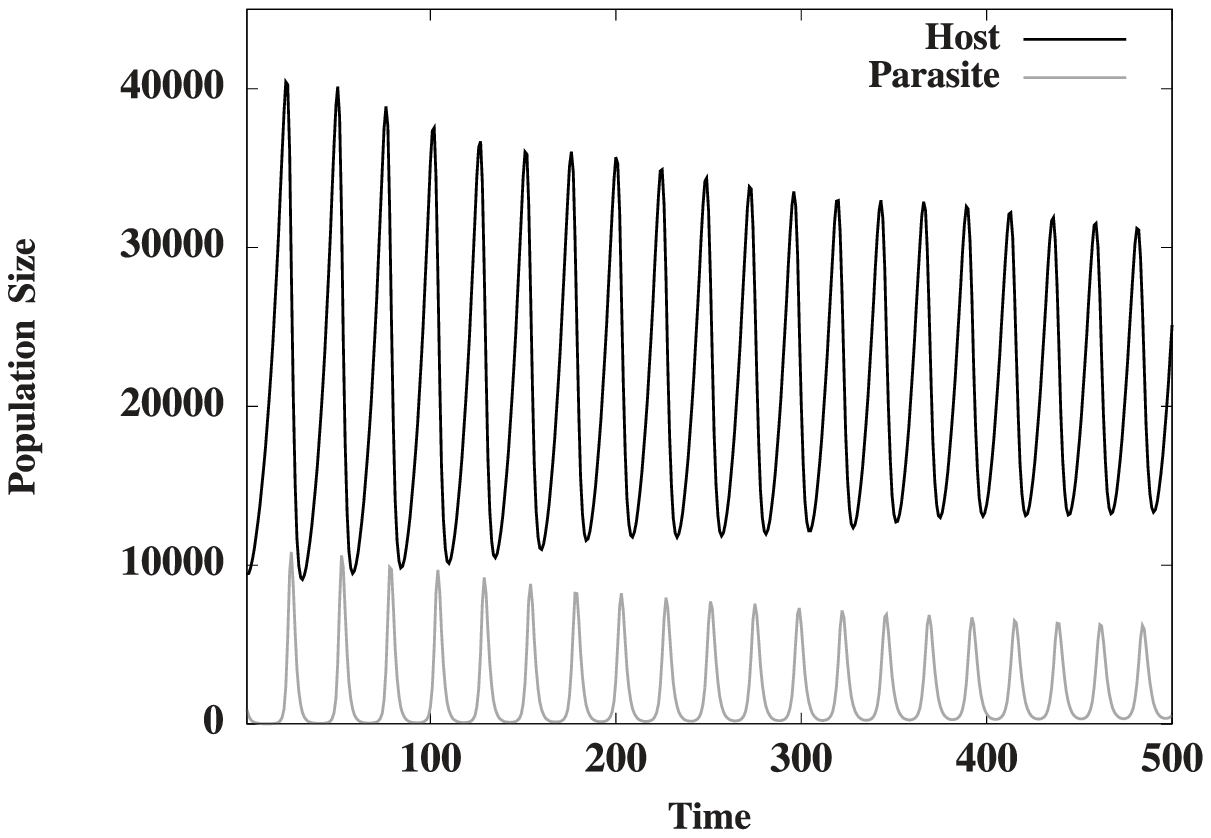}}\quad
	\caption{: The deterministic coevolutionary model with epidemiology of equation \eqref{WSAT_eq3} generates (a - top left) host and parasite populations fluctuations in time \index{demography! host-parasite co-demography}. (b - top right) Host and parasite diversity fluctuate in time measured as the relative $\pi$, and as (c - bottom left) the relative frequency of the singleton class over time (sample size of 20). In (b) and (c), we show the analytical expectations based on the Wright-Fisher model without overlap, and the stochastic simulations taking into account overlap in host and parasite generations due to disease transmission. (d - bottom right) Introducing a small amount of stochasticity for birth and death rates around the deterministic expectations of equation \eqref{WSAT_eq3}. The GFG model of Table \ref{WSAT_infMatrices} is run with the following parameters: $b=1$, $d=0.9$, $\beta= 0.00005$, $\delta=0.01$, $s=1$, $c_{H_1}=c_{P_2}=0.05$, and $c_{H_2}=c_{P_1}=0$.}
	\label{WSAT_Fig2}
\end{figure}

\indent Our first results show that the changes in population sizes are observable in the parasite population using the changes in SFS over time \cite{WSAT_zivkovic2019}. In Figure \ref{WSAT_Fig2}, this is done by tracking the number of singletons of the SFS or the pairwise diversity measure $\Pi$ over time. For the host, the change in SFS is very small and probably unobservable in data. In other words, using time sample data of the parasite population one can track the changes in population size due to the coevolutionary dynamics, and this signature is unique to a given combination of parameters and period/amplitude of cycles under the GFG model \cite{WSAT_zivkovic2019}. The reason that we can see the signature in parasites but not in hosts is that their time scales of evolution (of the coalescent process) are actually different. The time scale is defined by the initial size at the onset of the coevolutionary history ($N_{H}$=200,000 and $N_{P}$=20,000 in Figure 2). In biologically realistic cases, the initial size of the host is larger than that of the parasite as we assume that an epidemics starts from a small parasite population introduced into a mostly susceptible host population. However, dynamics under different interaction matrices such as GFG or MA models could not be distinguished based on SFS changes alone \cite{WSAT_zivkovic2019}. The results in Figure \ref{WSAT_Fig2} are obtained for one parasite generation per host generation and one parasite per host. When relaxing these assumptions, we can show that increasing the number of parasite generations per host does increase the possibility to observe the changes in population size using the parasite SFS, while increasing the number of parasite individuals per host has the opposite effect. Note that our approach can be extended to compute a sequential Markov coalescent (SMC) for host and parasite genomes, which takes recombination into account \cite{WSAT_dutheil2019, WSAT_sellinger2020}.

An interesting feature of our model 3 is the overlap of host generations implicitly assumed in epidemiological models for disease transmission. However, this assumption violates the Wright-Fisher model of non-overlapping generations that we apply to predict polymorphism data. \cite{WSAT_ZivkovicandStephan2011} \index{genetic drift! Wright-Fisher model} We thus built an ad hoc simulation model to test whether this deviation from the WF does influence our result. Two compartments are created at each host generation. A proportion $d$ of host die and are replaced by offspring produced at the previous generation. A proportion $1-d$ of host do survive to the next generation and are sampled without replacement from the previous generation. If $d=1$, we obtain the WF model, while if $d=1/N$ we obtain the discrete-time Moran model. Our results regarding the SFS obtained under model 3 are thus valid for values of $d$ close to 1 (Fig. \ref{WSAT_Fig2}b and c). Larger generation overlap (small $d$ values) only delays the time to reach the equilibrium state, as is apparent in the time scale difference between the Moran and the WF models. However, in our case of permanent non-equilibrium population size dynamics, the generation overlap generates a difference in $\Pi$ and SFS compared to the WF model which is not compensated over time \cite{WSAT_zivkovic2019}. \index{genetic drift! Cannings model} In biological systems where strong overlap of generations occurs, such as for long lived trees, caution should thus be applied when using the predictions from our coevolutionary model and SFS computations based on the WF model. Furthermore, note that the behaviour of model 3 can be affected by small stochastic perturbations in birth and death processes, so that cycles of population size change may dampen off (Fig. \ref{WSAT_Fig2}d).
\\ 
\indent Current work in progress includes the possibility in model 3 that the parasite population deviates from the WF model. In fact, population models with large variance in offspring production are suggested to be realistic for many plant pathogens \cite{WSAT_tellier2014}. Backward in time, the underlying population genealogy would follow the so-called multiple merger class of coalescent models (\textit{e.g.} $\Lambda$-coalescent or $\beta$-coalescent) \index{coalescent! $\Lambda$-coalescent} \index{coalescent! $\beta$-coalescent} which have been studied in depth in other projects of this SPP \cite{WSAT_birkner2019, WSAT_freund2019}. The forward computation in \cite{WSAT_ZivkovicandStephan2011} of the SFS for variable population size need to be updated to include a Fleming-Viot process and large variance in offspring number. \index{genetic drift! Cannings model}

\subsection{Current work in progress: stochastic disease transmission}
A last intrinsic factor of stochasticity lies in the process of disease transmission itself, as diseased individuals' rate of contact is stochastic and can deviate from the mass action assumption of model 3 (and classic epidemiological models, \cite{WSAT_MayandAnderson1983,WSAT_AshbyBoots2017}). We have simplified model 3 (MA model, $c_{H}=c_{P}=s=0, d=b=1$, equation \eqref{WSAT_eq4}) so that the total population size is fixed and constant:
\begin{eqnarray}
\label{WSAT_eq4}
\frac{dH_i}{dt}&=&-H_i\beta\sum\limits_{j=1}^{A}\alpha_{ij}\sum\limits_{k=1}^{A}I_{kj},\\
\frac{dI_{ij}}{dt}&=&I_{ij}(-1-\delta_{ij})+H_i\beta\alpha_{ij}\sum\limits_{k=1}^{A}I_{kj} \nonumber.
\end{eqnarray}

Variance in disease transmission has been investigated in \cite{WSAT_allen2008} via a discrete time Markov chain (DTMC) epidemic model. \index{Markov chain! discrete time} We chose a time step $\Delta t$ sufficiently small such that the number of infected individuals changes by at most one over all infected types ($I_{ij},\forall i,j$) during the interval: $I_{ij} \rightarrow I_{ij}+1$, $I_{ij} \rightarrow I_{ij}-1$, or $I_{ij} \rightarrow I_{ij}$. At each time step $\Delta t$, the transition probabilities for the DTMC epidemic model for a given $I_{ij}$ are:
\begin{equation}
\label{WSAT_eq5}
P_{I_{ij}(t)\rightarrow I_{ij}(t+\Delta t)}=\left 
\{
\begin{array}{@{}ll@{}}
\beta H_{i}\alpha_{ij}\sum\limits_{k=1}^{A}I_{kj} \Delta t,\,\,\, \text{for } I_{ij}(t+\Delta t)=I_{ij}(t)+1 \\
(1+\delta_{ij})I_{ij} \Delta t,\,\,\, \text{for } I_{ij}(t+\Delta t)=I_{ij}(t)-1 \\
1-[\beta H_{i}\alpha_{ij}\sum\limits_{k=1}^{A}I_{kj} \Delta t + (1+\delta_{ij})I_{ij} \Delta t],\,\,\, \\
\text{for } I_{ij}(t+\Delta t)=I_{ij}(t). \\
\end{array}\right.
\end{equation}

In principle one obtains a quasi-equilibrium density of infected individuals (of type $ij$), that is stochastic variation at the vicinity of the deterministic equilibrium density, when assuming that the disease transmission function (defined by the product $\beta \alpha_{ij}H_{i}I_{ij}$) follows a normal distribution with the number of infected individuals (${I_{ij}}$) as mean. However, analytical solutions are cumbersome to obtain if we assume cycling population size and allele frequencies, and we resort to simulations for this part of our study. The stochastic behaviour chosen in \cite{WSAT_allen2008} has a relatively weak effect on the outcome of the model (stability of the polymorphic equilibrium point). If larger variance of disease transmission is assumed, the variance and stability of the quasi-equilibrium are expected to be strongly affected. Moreover, a further question will be to assess the influence of other stochastic processes described above in nudging the population away from the quasi-equilibrium. We speculate that genetic drift will have a strong effect in moving the population towards the disease extinction or fully diseased host population state. Finally, the stability of the dynamical system and the maintenance of polymorphism under our model 3 (Eq~\ref{WSAT_eq4}) can be also studied under the evolution of the system's coefficients due to the introduction of host and parasite mutants with different strategies (\textit{e.g.} for $\alpha_{ij}$, $\beta_{ij}$, ...). Different evolutionary stable strategies exist for the host and parasite coefficients of the infection matrix $\alpha_{ij}$ depending on the chosen shapes of the trade-off with the costs $c_{H_i}$ and $c_{P_j}$. The work in \cite{WSAT_best2017} can be extended to include stochasticity following the principles reviewed in \cite{WSAT_bovier2019} on dynamical systems evolving under adaptive dynamics. \index{adaptive dynamics}

\section{Extrinsic stochasticity}

\subsection{Spatial structure and migration}
In the previous part we have considered intrinsic factors of stochasticity influencing coevolutionary dynamics and simulated expected signatures of coevolution at the underlying loci and across whole genomes. However, most coevolutionary processes do happen in spatially structured populations of hosts and parasites linked by gene flow. Previous studies have investigated deterministically the coevolutionary dynamics in spatially homogeneous populations, \textit{i.e.} with identical values of the coevolutionary parameters across populations. A spatial model 1 with genetic drift in each population and stochastic gene flow does show fixation of alleles \cite{WSAT_moreno2013}. However, we have obtained a stable deterministic dynamics in a heterogeneous spatial model based on model 1 with two populations of host and parasite linked by gene flow. Stable deterministic dynamics can arise due to ndFDS when one or several model parameters differ in values between populations such as the costs $c_{H}$ and $c_{P}$ or the disease prevalence $\phi$ (\textit{e.g.} \cite{WSAT_moreno2013}). In this case, a small simulation study shows that including genetic drift and stochastic gene flow between the two populations still generates trench warfare dynamics. However, a more complex model including a multi-locus GFG model with a 2D-spatial structure, stochastic gene flow, genetic drift within populations, and mutation considerably slowed down the time for allele fixation \cite{WSAT_thrall2002}. Combining several sources of stochasticity can therefore increase the time until genotypes are fixed and thus generates a stochastic polymorphism behaviour \cite{WSAT_allen1975}.

\subsection{Current work in progress: environmental stochastic variation in parameters of coevolution}
Another possibly important source of variability is stochasticity associated with environmental variables. For many parasites of plants and invertebrates, the success of infection depends indeed highly on environmental conditions. Therefore, variation between years or between seasons of climatic variables such as humidity or temperature can have a great influence on the dynamics of epidemics. In our model 1 for example, this would mean to consider $\phi$ as a random temporal variable. The cost of infection $s$ could also be considered as a random temporal variable. Regarding model 3, one could consider the disease transmission parameter $\beta$, the virulence $\delta$ or the cost of infection $s$ to be also random temporal variables. The largest effects of stochasticity on dynamics are  expected to be due to $\beta$ as it strongly determines the type and outcome of coevolutionary dynamics \cite{WSAT_zivkovic2019}. This aspect of stochasticity has been largely ignored in the literature so far, while being potentially important for many infectious diseases.

\subsection{Changes in population sizes or allele frequencies due to higher-order interactions}
A last source of stochasticity deals with the effect of higher-order interactions on hosts or their parasites from other trophic levels. In other words, hosts can be also attacked by several parasite species and/or can be prey of some predator species. Similarly, parasites (fungi, bacteria) can themselves be infected by other parasites (viruses, phages). Another possibility includes parasites with complex life cycles, as they require to infect several host species to complete their cycle. At each stage of infection, intrinsic and extrinsic stochastic processes do occur. As far as we know, this type of stochastic process has also been ignored in the literature, while multi-layer (nested) stochastic processes have received recent attention in mathematics \cite{WSAT_dawson2003, WSAT_birkner2019, WSAT_sturmwinter2019}.

\section{Conclusion}
Several stochastic processes do occur in host and parasite populations and can affect and change the outcome of the coevolutionary dynamics compared to the widely studied deterministic models. Therefore, several of our studies aimed at investigating the effects of genetic drift and mutation at the locus under coevolution but also the effect of co-demographic history on genome-wide polymorphism of interacting species. These results have been instrumental in developing the first inference method to estimate coevolutionary parameters based on host and parasite polymorphism data \cite{WSAT_maerkle2019}. Extension of this work includes taking into account possible large variances in offspring production. Moreover we highlight here the lack of study of extrinsic stochastic effects on coevolutionary outcome, which may be very prevalent in natural systems in contrast to controlled laboratory experiments.

\section{Acknowledgements}
We are indebted to Daniel \v{Z}ivkovi{\'c}, Sona John, Hanna M\"{a}rkle and Stephanie Moreno-Gamez who contributed to the analysis of the coevolutionary models. AT acknowledges numerous discussions over the years with James Brown, Jerome Enjalbert and Tatiana Giraud about plant-parasite coevolution. AT is also funded in part by the DFG grant TE809/3-1 and TE809/4-1 from the Priority Program 1819 on Rapid Evolutionary Adaptation.




\end{document}